\documentstyle[12pt]{article}
\renewcommand{\baselinestretch}{1.47}

\begin{document}
\parskip=5pt plus 1pt minus 1pt

\begin{flushright}
{\bf BIHEP-TH-95-2}\\
{\bf LMU-02/95}\\
{\large January 1995}
\end{flushright}

\vspace{0.2cm}

\begin{center}
{\Large\bf Effects of the Space-like Penguin Diagrams}\\
{\Large\bf on $CP$ Asymmetries in Exclusive $B$ Decays}
\end{center}

\vspace{0.4cm}

\begin{center}
{\bf Dongsheng DU} \\
{\sl CCAST (World Laboratory), P.O. Box 8730, Beijing 100080, China} \\
{\sl Theory Division, Institute of High Energy Physics, Academia Sinica,} \\
{\sl P.O. Box 918 (4), Beijing 100039, China}
\end{center}

\begin{center}
{\bf Zhi-zhong XING}\footnote{Alexander von Humboldt Research Fellow}\\
{\sl Sektion Physik, Theoretische Physik, Universit$\ddot{a}$t
M$\ddot{u}$nchen,}\\
{\sl Theresienstrasse 37, D-80333 Munich, Germany}
\end{center}

\vspace{1.6cm}

\begin{abstract}

The space-like penguin contributions were ignored on little ground in many
previous studies of nonleptonic
$B$ transitions and $CP$ violation. Taking the penguin-dominated channels
$B^{-}_{u}\rightarrow \bar{K}^{0}\pi^{-}$ and $B^{-}_{u}\rightarrow K^{0}K^{-}$
for
example, we illustrate the non-negligible effects of the space-like penguin
diagrams
on $CP$ asymmetries. Some qualitative remarks are given on the gluonic penguin
picture and
direct $CP$ violation in exclusive two-body decays of $B$ mesons.

\end{abstract}

\newpage

Within the standard electroweak model, direct $CP$ violation is expected to
manifest
itself significantly in some exclusive $B$-meson decays which are dominated by
the
one-loop gluonic penguin transitions $b\rightarrow qg^{*}\rightarrow
q(q^{~}_{g}\bar{q}^{~}_{g})$
($q=d$ or $s$ and $q^{~}_{g}=u,d$, or $s$) [1-4]. Non-zero $CP$ asymmetries can
arise
through the interference between two independent amplitudes that have both
different
$CP$ violating phases and different $CP$ conserving phases. In the time-like
penguin
diagram, the necessary strong phases are provided by different loop effects
of internal $u$ and $c$ quarks involving real (on-shell) particle rescattering.
In contrast, the space-like penguin diagram can only provide an overall
$CP$ conserving phase due to final state hadronization. Thus the time-like
penguin
amplitudes play the key role in giving rise to $CP$ violation, while the
space-like
penguin amplitudes only take effects by modifying the dispersive or absorptive
parts of the time-like ones. In many previous studies of $CP$ violation in $B$
decays,
the contributions from those annihilation-type diagrams were neglected for the
argument that
they should be formfactor suppressed or helicity unfavoured. This argument
becomes
questionable, however, for the space-like penguin channels since their
amplitudes
can be remarkably enhanced by the hadronic matrix elements involving
$(V-A)(V+A)$ or $(S-P)(S+P)$
currents [5]. To date, a phenomenological demonstration of the space-like
penguin effects
on $CP$ asymmetries in exclusive $B$ decays has been lacking.

In this work we shall take the penguin-dominated decay modes
$B^{-}_{u}\rightarrow
\bar{K}^{0}\pi^{-}$ and $K^{0}K^{-}$ for example to illustrate the
non-negligible
contributions of the space-like penguin amplitudes to $CP$ asymmetries. A
simple
kinematic picture is presented for decays of the type $B\rightarrow PP$ in
order
to minimize the uncertainty with the gluonic momentum transfer. We calculate
decay
amplitudes by using the effective weak Hamiltonian and factorization
approximation.
Our numerical results show that the space-like penguin diagrams can
significantly affect
the $CP$ violating signals in exclusive $B$ transitions. It is therefore
worthwhile to
reexamine some previous works on the penguin-dominated $B$ decays and $CP$
violation,
in which the space-like penguin effects were ignored on little ground.

The transitions $B^{-}_{u}\rightarrow \bar{K}^{0}\pi^{-}$ and
$B^{-}_{u}\rightarrow
K^{0}K^{-}$ are of great interest for studying direct $CP$ violation [1-4] and
extracting the Kobayashi-Maskawa (KM) phase parameters [6]. They
have two advantages for our present purpose. First, they occur only through the
tree-level annihilation and penguin channels in the quark-diagram scheme [5,7].
The former can be safely neglected in comparison with the space-like penguin
contribution. Second, the electroweak penguin effects on these two decay modes
are
rather small [8], thus one may simply apply the QCD-loop induced Hamiltonian to
them.
The one-loop gluonic penguin Hamiltonian for $B$ decays is given by [1-5]
\begin{equation}
{\cal H}_{\rm eff}(\Delta B=-1)\;=\; -\frac{G_{\rm F}}{\sqrt{2}}
\frac{\alpha_{s}}{8\pi} \left [\sum_{i}v_{i}\left (F^{\rm T}_{i}+F^{\rm
S}_{i}\right )\right ]
\left (-\frac{Q_{3}}{N_{c}}+Q_{4}-\frac{Q_{5}}{N_{c}}+Q_{6}\right ) \; ,
%		(1)
\end{equation}
where QCD corrections are approximately included by the
effective coupling constant $\alpha_{s}$ at the physical scale
$\mu =m_{b}$; $v_{i} = V_{ib}V^{*}_{iq}$ ($i=u,c,t$
and $q=d,s$) are the KM factors corresponding to
$b\rightarrow q$; $F^{\rm T}_{i}$ and $F^{\rm S}_{i}$ stand for the loop
integral functions of
the time-like (T) and space-like (S) penguin diagrams respectively;
$N_{c}$ is the number of colors; and $Q_{3,...,6}$ represent gluonic penguin
operators of the
form [9]
\begin{equation}
\begin{array}{lll}
& Q_{3} & =\; (\bar{q}b)^{~}_{V-A}
\displaystyle\sum_{q^{~}_{g}}(\bar{q}^{~}_{g}q^{~}_{g})^{~}_{V-A}\; ,\;\;\;\;\;
Q_{4}\; =\; (\bar{q}^{\alpha}b^{\beta})^{~}_{V-A}
\displaystyle\sum_{q^{~}_{g}}(\bar{q}^{\beta}_{g}q^{\alpha}_{g})^{~}_{V-A}\;
,\\
& Q_{5} & =\; (\bar{q}b)^{~}_{V-A}
\displaystyle\sum_{q^{~}_{g}}(\bar{q}^{~}_{g}q^{~}_{g})^{~}_{V+A}\; ,\;\;\;\;\;
Q_{6}\; =\; (\bar{q}^{\alpha}b^{\beta})^{~}_{V-A}
\displaystyle\sum_{q^{~}_{g}}(\bar{q}^{\beta}_{g}q^{\alpha}_{g})^{~}_{V+A}\;
.\nonumber
%		(2)
\end{array}
\end{equation}
Note that $F^{\rm T,S}_{i}$ are dependent on the gluonic momentum transfer
$k^{2}$.
Without loss of generality, we obtain the analytical expressions of $F^{\rm
T}_{i}$ and $F^{\rm S}_{i}$ as [10]
\begin{eqnarray}
F^{\rm S}_{i=u,c} & = & \displaystyle \frac{10}{9}-\frac{2}{3}\ln a_{i} +
\frac{2}{3}r_{i}
- \frac{1}{3}(2+r_{i})\sqrt{1-r_{i}} \left [ \ln \left | 1+\sqrt{1-r_{i}}\right
|
- \ln \left | 1-\sqrt{1-r_{i}}\right |\right ] \; , \nonumber \\
F^{\rm T}_{i=u,c} & = & \displaystyle \frac{10}{9}-\frac{2}{3}\ln a_{i} +
\frac{2}{3}r_{i}
- \frac{1}{3}(2+r_{i})\sqrt{|1-r_{i}|}\left [ 2\; {\rm arccot}\sqrt{r_{i}-1}
{}~\theta (r_{i}-1)
\right . \nonumber \\
&  & \left . + \left (\ln \left |1+\sqrt{1-r_{i}}\right | - \ln \left
|1-\sqrt{1-r_{i}}\right |
-{\rm i}\pi\right ) \theta (1-r_{i}) \right ] \; , \\
F^{\rm S}_{i=t} & = & F^{\rm T}_{i=t} \; =\; \displaystyle
\frac{(18-11a_{i}-a^{2}_{i})a_{i}}
{12(1-a_{i})^{3}} - \frac{4-16a_{i}+9a^{2}_{i}}{6(1-a_{i})^{4}}\ln a_{i} \; ,
\nonumber
%		(3)
\end{eqnarray}
where $a_{i}=m^{2}_{i}/m^{2}_{W}$ , $r_{i}=4m^{2}_{i}/k^{2}$, and $F^{\rm
S}_{i=t}$ = $F^{\rm T}_{i=t}$
= $43/72$ at the limit $m_{t}=m^{~}_{W}$ .

Between a decay mode $B^{-}_{u}\rightarrow f$ and its $CP$-conjugate
counterpart $B^{+}_{u}\rightarrow \bar{f}$,
the $CP$ asymmetry ${\cal A}_{f}$ is defined as the ratio of the difference to
the sum of
their decay rates. Since $v_{u}+v_{c}+v_{t}=0$, each decay amplitude can be
decomposed into two
terms which are proportional to $v_{u}$ and $v_{c}$ respectively.
Hence ${\cal A}_{f}$ depends only upon the KM phase $(v_{u}v^{*}_{c})$ and the
quantity
\begin{equation}
R_{f} \; =\; \frac{\left (F^{\rm T}_{c}-F^{\rm T}_{t}\right )+\xi_{f} \left
(F^{\rm S}_{c}-F^{\rm S}_{t}\right )}
{\displaystyle\left (F^{\rm T}_{u}-F^{\rm T}_{t}\right ) +\xi_{f}
\displaystyle\left (F^{\rm S}_{u}-F^{\rm S}_{t}\right )} \; ,
%		(4)
\end{equation}
consisting of the non-trivial strong phases due to final state interactions.
Here
the parameter $\xi_{f}$ measures the relative size and sign between the
space-like and
time-like penguin amplitudes. In many previous studies $\xi_{f}=0$ was assumed.
Explicitly the $CP$ asymmetry ${\cal A}_{f}$ can be expressed as
\begin{equation}
{\cal A}_{f} \;=\; \frac{2{\rm Im}(v_{u}v^{*}_{c})\cdot {\rm Im}R_{f}}
{|v_{u}|^{2}+|v_{c}|^{2}\cdot |R_{f}|^{2}+2{\rm Re}(v_{u}v^{*}_{c})
\cdot {\rm Re}R_{f}}\; .
%		(5)
\end{equation}
In the following we shall discuss how to evaluate the loop intergral functions
$F^{\rm T,S}_{i}$ and the hadronization parameter $\xi_{f}$ in order to
quantitatively determine
$R_{f}$ and ${\cal A}_{f}$ .

The problem with $F^{\rm T}_{i}$ and $F^{\rm S}_{i}$ is the unknown value of
$k^{2}$, the four-momentum squared of the
virtual gluon in the exclusive penguin channels. For the time-like penguin
transitions, one used to pick a special
value from $k^{2}\in \left (0, m^{2}_{b}\right ]$ or $k^{2}\in \left
[m^{2}_{b}/4, m^{2}_{b}/2 \right ]$ with some
kinematic arguments [4]\footnote{Considering an additional hard gluon to
accelerate the
spectator quark in a time-like penguin graph, Simma and Wyler [1] have
calculated the $k^{2}$ distribution
of some charmless exclusive $B$ decays and folded it with the momentum
dependence of the loop
amplitudes. The relevant branching ratios yielded in this method are however
smaller than those
from other ways.}.
Taking the space-like penguin diagrams into account, here we present a
simple kinematic picture for two-body penguin-induced decays $B\rightarrow XY$
as illustrated in Fig. 1. In the rest frame of the $B$ meson, we assume: (a)
the spectator quark of the time-like penguin graph
has negligibly small momentum in either the initial or the final state; (b) the
two quarks forming the
meson $X$ in Fig. 1(a) have the same momentum; and (c) the momentum of the
quark pair created from the vacuum
are negligible in the space-like penguin graph Fig. 1(b). Accordingly we find
that the average value of the
gluonic momentum transfer $k^{2}$ can be given by
\begin{equation}
\langle k^{2}\rangle \; =\; m^{2}_{b}+m^{2}_{q}-2m_{b}E_{q} \; ,
%		(6)
\end{equation}
where $E_{q}$ is determinable from
$$
E_{q}+\sqrt{E^{2}_{q}-m^{2}_{q}+m^{2}_{q^{~}_{g}}}+\sqrt{4E^{2}_{q}-4m^{2}_{q}+m^{2}_{q^{~}_{g}}} \; =\; m_{b} \;
\eqno(7{\rm a})
$$
for the time-like penguin channels; or from
$$
E_{q}+\sqrt{E^{2}_{q}-m^{2}_{q}+m^{2}_{q^{~}_{g}}} \; =\; m_{b}+m_{q^{~}_{g}}
\;
\eqno(7{\rm b})
$$
for the space-like penguin channels.
In the case of $q=q^{~}_{g}$ , we obtain $\langle k^{2}\rangle_{\rm
T}=\frac{1}{2}\left (m^{2}_{b}-m^{2}_{q}\right )>0$
and $\langle k^{2}\rangle_{\rm S}=m_{q}(m_{q}-m_{b})<0$. From Eq. (7)
we observe that taking $|\langle k^{2}\rangle_{\rm T}| = |\langle
k^{2}\rangle_{\rm S}|$ is in general not true for
phenomenology.

The above kinematic picture, based on the valence-quark assumption, has avoided
the drawback
of taking arbitrary values for $k^{2}$ in the studies of exclusive
penguin-mediated $B$ decays.
However, such an analysis cannot reflect the dynamics of final state
hadronization. For example,
it predicts the same size of $\langle k^{2}\rangle_{\rm T}$ (or $\langle
k^{2}\rangle_{\rm S}$)
for the processes $B^{-}_{u}\rightarrow K^{(*)-} + (\pi^{0}, \eta^{0},
\rho^{0})$.
Hence it is problematic to apply this approach to all two-body $B$ transitions.
Pursuing a phenomenological insight into the time-like and space-like penguin
diagrams,
we believe that the simple picture given above should be applicable to
those $B$ decays into two pseudoscalar mesons.
Whether parallel estimates of $\langle k^{2}\rangle$ can be carried out for
penguin-induced $B$
decays of the types $B\rightarrow PV$ and $B\rightarrow VV$ is still an open
question.

Applying the effective Hamiltonian ${\cal H}_{\rm eff}$ and factorization
approximation [11] to
$B^{-}_{u}\rightarrow$ $\bar{K}^{0}\pi^{-}$ and $K^{0}K^{-}$, one obtains
\setcounter{equation}{7}
\begin{equation}
\begin{array}{lll}
\xi_{\bar{K}^{0}\pi^{-}} & = & \displaystyle \left [
1+\frac{2m^{2}_{B^{-}_{u}}}{(m_{s}-m_{u})(m_{b}+m_{u})}\right ]
\cdot \left [ 1+\frac{2m^{2}_{\bar{K}^{0}}}{(m_{s}+m_{d})(m_{b}-m_{d})}\right
]^{-1}\cdot Z_{\bar{K}^{0}\pi^{-}} \; , \\
\xi_{K^{0}K^{-}} & = & \displaystyle \left [
1+\frac{2m^{2}_{B^{-}_{u}}}{(m_{d}-m_{u})(m_{b}+m_{u})}\right ]
\cdot \left [ 1+\frac{2m^{2}_{K^{0}}}{(m_{d}+m_{s})(m_{b}-m_{s})}\right ]^{-1}
\cdot Z_{K^{0}K^{-}} \; .
%		(8)
\end{array}
\end{equation}
In this equation, the terms proportional to $m^{2}_{B^{-}_{u}}$
($m^{2}_{\bar{K}^{0}}$ or
$m^{2}_{K^{0}}$) arise from transforming the $(V-A)(V+A)$ currents into the
$(V-A)(V-A)$ ones
for the space-like (time-like) penguin amplitudes; and the parameter $Z_{f}$
($f=\bar{K}^{0}\pi^{-}$
or $K^{0}K^{-}$) describes the ratio of the space-like hadronic matrix element
to the time-like
one after factorization. In terms of decay constants and formfactors [12,5],
$Z_{\bar{K}^{0}\pi^{-}}$
and $Z_{K^{0}K^{-}}$ can be expressed as
\small
\begin{equation}
\begin{array}{lll}
Z_{\bar{K}^{0}\pi^{-}} & = & -
\displaystyle\frac{m^{~}_{\bar{K}^{0}}-m_{\pi^{-}}}{m^{~}_{\bar{K}^{0}}+m_{\pi^{-}}}
\cdot \frac{m_{B^{-}_{u}}+m_{\pi^{-}}}{m_{B^{-}_{u}}-m_{\pi^{-}}}\cdot
\frac{(m^{~}_{\bar{K}^{0}}+m_{\pi^{-}})^{2}-m^{2}_{B^{-}_{u}}}{(m_{B^{-}_{u}}+m_{\pi^{-}})^{2}
-m^{2}_{\bar{K}^{0}}}\cdot \frac{f_{B^{-}_{u}}F^{\rm a}_{+}(m^{2}_{B^{-}_{u}})}
{f_{\bar{K}^{0}}F^{B^{-}_{u}\pi^{-}}_{+}(m^{2}_{\bar{K}^{0}})} \; , \\
Z_{K^{0}K^{-}} & = & -
\displaystyle\frac{m^{~}_{K^{0}}-m^{~}_{K^{-}}}{m^{~}_{K^{0}}+m^{~}_{K^{-}}}
\cdot \frac{m_{B^{-}_{u}}+m^{~}_{K^{-}}}{m_{B^{-}_{u}}-m^{~}_{K^{-}}}\cdot
\frac{(m^{~}_{K^{0}}+m^{~}_{K^{-}})^{2}-m^{2}_{B^{-}_{u}}}{(m_{B^{-}_{u}}+m^{~}_{K^{-}})^{2}
-m^{2}_{K^{0}}}\cdot \frac{f_{B^{-}_{u}}F^{\rm a}_{+}(m^{2}_{B^{-}_{u}})}
{f_{K^{0}}F^{B^{-}_{u}K^{-}}_{+}(m^{2}_{K^{0}})} \; ,
%		(9)
\end{array}
\end{equation}
\normalsize
where the annihilation formfactor $F^{\rm a}_{+}(m^{2}_{B^{-}_{u}}) ={\rm
i}16\pi \alpha_{s}
f^{2}_{B^{-}_{u}}/m^{2}_{B^{-}_{u}}$ is given by QCD calculations [13]. It
should be noted that $Z_{f}$ or
$\xi_{f}$ is primarily absorptive and is not negligible. The
presence of the space-like penguin amplitudes can correct ${\rm Im}F^{\rm
T}_{i}$ through
Eq. (4). In contrast, the perturbative QCD corrections mainly modify ${\rm
Re}F^{\rm T}_{i}$, as
shown in Ref. [3]. Thus evaluating the size of $Z_{f}$ is very desirable in
order to
properly calculate $CP$ asymmetries in the penguin-mediated $B$ decays.

For illustration, let us estimate the $CP$ asymmetry ${\cal A}_{f}$
($f=\bar{K}^{0}\pi^{-}$ or
$K^{0}K^{-}$) and compare the result of $\xi_{f}\neq 0$ with that of
$\xi_{f}=0$.
We use the current quark masses $(m_{u}, m_{d}, m_{s}, m_{c}, m_{b}, m_{t})$ =
(0.005, 0.01, 0.175,
1.35, 4.8, 170) in unit of GeV and the meson masses $(m_{\pi^{-}},
m^{~}_{K^{0}}, m^{~}_{K^{-}}, m_{B^{-}_{u}})$ =
(139.6, 497.7, 493.7, 5278.7) in unit of MeV [14]. The decay constants and
formfactors
are taken as $f_{\pi^{-}}=130.7$ MeV, $f_{K^{0}}=f_{\bar{K}^{0}}\approx
f_{K^{-}}$ = 159.8 MeV [14],
$f_{B^{-}_{u}}\approx 1.5f_{\pi^{-}}$ [15]; $F^{B^{-}_{u}\pi^{-}}_{+}(0)\approx
0.29$, and
$F^{B^{-}_{u}K^{-}}_{+}(0)\approx 0.32$ [16]. The inputs of the Wolfenstein
parameters [17]
are $A=0.80$, $\lambda=0.22$, $\rho=-0.07$, and $\eta=0.38$ [18]. After a
straightforward
calculation we find $\xi_{\bar{K}^{0}\pi^{-}}\approx$ i1.3, ${\cal
A}_{\bar{K}^{0}\pi^{-}}\approx
1.3\%$ and $\xi_{K^{0}K^{-}}\approx$ i0.32, ${\cal A}_{K^{0}K^{-}}\approx
-10.9\%$. In
comparison with the values ${\cal A}_{\bar{K}^{0}\pi^{-}}\approx 0.4\%$ and
${\cal A}_{K^{0}K^{-}}\approx -6.4\%$, which are obtained by taking
$\xi_{\bar{K}^{0}\pi^{-}}$
= $\xi_{K^{0}K^{-}}$ =0, we observe that contributions from the space-like
penguin amplitudes
can significantly enhance the $CP$ asymmetries.

More generally one may assume an arbitrary phase shift $\theta_{f}^{~}$ between
the space-like
and time-like penguin amplitudes. Replacing $\xi_{f}$ by $|\xi_{f}|e^{{\rm
i}\theta^{~}_{f}}$ in Eq. (4),
we examine the dependence of ${\cal A}_{f}$ upon $\theta^{~}_{f}$ for the decay
modes
$B^{-}_{u}\rightarrow \bar{K}^{0}\pi^{-}$ and $K^{0}K^{-}$. The numerical
results are
shown in Fig. 2, where $\theta^{~}_{\bar{K}^{0}\pi^{-}}$ and
$\theta^{~}_{K^{0}K^{-}}$ change
from $-180^{0}$ to $180^{0}$. It is clear that the space-like penguin
transitions play
an important role for $CP$ violation in these two processes.

Certainly the numbers given above have many uncertainties that are unable to be
removed to the
limit of our present understanding of nonleptonic weak decays and
nonperturbative confinement
forces [19]. While the quantitative results might not be trustworthy, we
emphasize that
qualitatively effects of the space-like penguin diagrams on $CP$ asymmetries
(and branching ratios)
in exclusive $B$ decays should be non-negligible. Whether such effects are
constructive or destructive
to $CP$ violating signals depends upon final state interactions of the decay
modes
under discussion. Our conclusion is that further studies of the
penguin-dominated $B$
decays (in particular, those promising processes for probing direct $CP$
violation or
testing unitarity of the KM matrix) are very necessary in order to advance our
understanding
of the ``penguin physics'' [20] and the underlying mechanism of direct $CP$
violation. \\

One of us (Z.Z.X.) would like to thank Professor H. Fritzsch for his warm
hospitality
and constant encouragement. He is also grateful to Professors W. S. Hou and A.
Khodjamirian
for useful communications.
This work was supported in part by the Alexander von Humboldt Foundation of
Germany
and by the National Nature Science Foundation of China.

\newpage

\begin{flushleft}
{\Large\bf Figure Captions}
\end{flushleft}

\vspace{1cm}

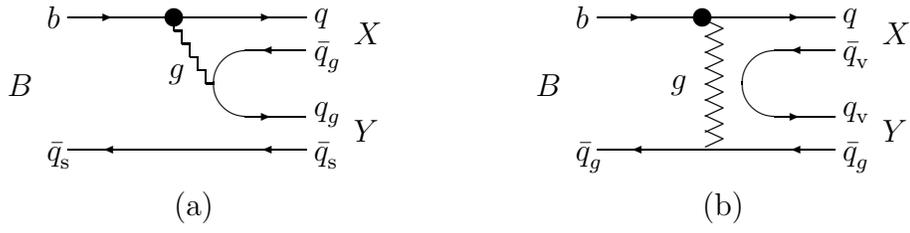
{\bf Figure 1:} $~$ Quark diagrams for a $B$ meson decaying into two light
mesons $X$ and $Y$
through the gluonic penguin process $b\rightarrow qg^{*}\rightarrow
q(q^{~}_{g}\bar{q}^{~}_{g})$:
(a) the time-like penguin, and (b) the space-like penguin. The dark dot stands
for the effective
four-fermion interactions $Q_{3}$ -- $Q_{6}$ . The subscripts ``s'' and ``v''
denote
``spectator'' and ``vacuum'', respectively. \\

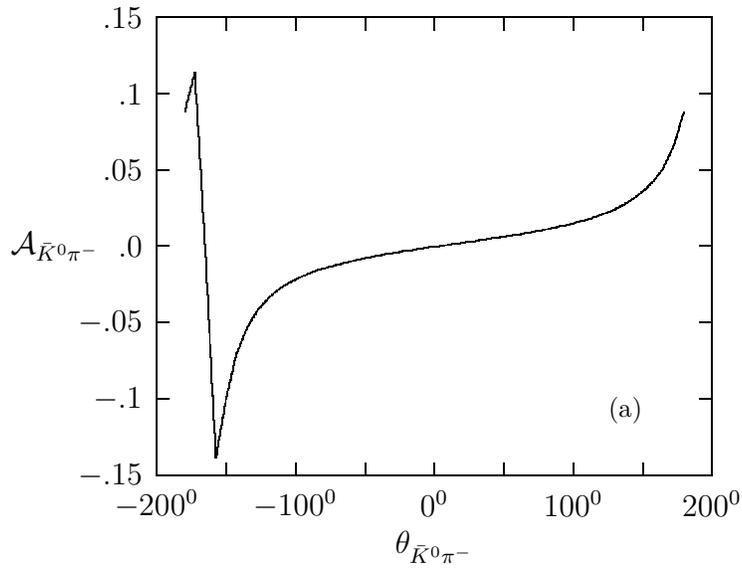
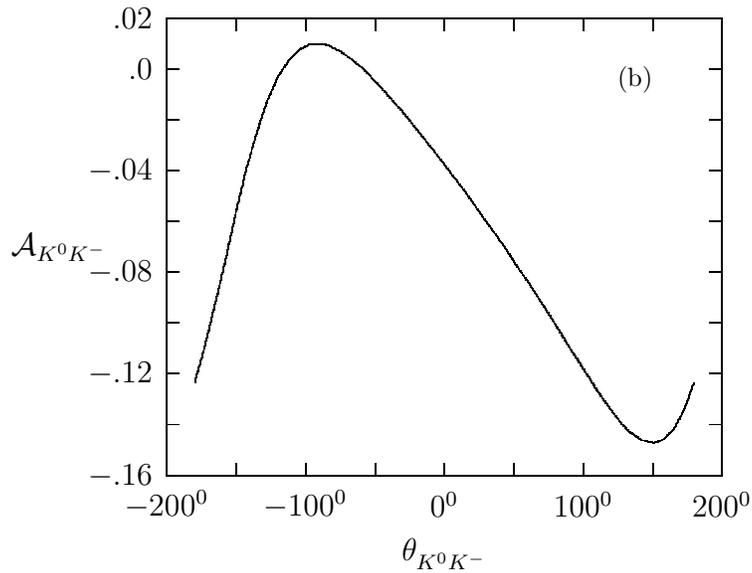
{\bf Figure 2:} $~$ Dependence of the $CP$ asymmetry ${\cal A}_{f}$ upon the
phase shift $\theta^{~}_{f}$
between the space-like and time-like penguin amplitudes: (a)
$B^{-}_{u}\rightarrow \bar{K}^{0}\pi^{-}$,
and (b) $B^{-}_{u}\rightarrow K^{0}K^{-}$.

\newpage

\begin{figure}

\vspace{-5.5cm}
\caption{}
%{Quark diagrams for a $B$ meson decaying into two light mesons $X$ and $Y$
%through the gluon-mediated process $b\rightarrow qg^{*}\rightarrow
%%q(q^{~}_{g}\bar{q}^{~}_{g})$:
%(a) the time-like penguin; (b) the space-like penguin. The dark dot stands for
%%the
%effective four-fermion interactions $Q_{3,...,6}$ .}
%\end{figure}

%\begin{figure}
% GNUPLOT: LaTeX picture
\setlength{\unitlength}{0.240900pt}
\ifx\plotpoint\undefined\newsavebox{\plotpoint}\fi
\sbox{\plotpoint}{\rule[-0.175pt]{0.350pt}{0.350pt}}%
%
\vspace{1cm}
\caption{}
%{The $CP$ asymmetry ${\cal A}_{f}$ versus an arbitrary phase shift $\theta$
%between the space-like and time-like penguin amplitudes in the decay mode {\bf
%%(a)} $B^{-}_{u}\rightarrow
%\bar{K}^{0}\pi^{-}$ and {\bf (b)} $B^{-}_{u}\rightarrow K^{0}K^{-}$. }
\end{figure}

\end{document}